\documentclass{IEEEtran}
\usepackage{listings}
\usepackage{times}
\usepackage{graphicx}
\usepackage[strings]{underscore}

\begin{document}

\title{ePython: An implementation of Python for the many-core Epiphany coprocessor}
\author{Nick Brown\\
       EPCC\\
       James Clerk Maxwell Building\\
       Edinburgh\\
       nick.brown@ed.ac.uk}


\maketitle

\begin{abstract}
The Epiphany is a many-core, low power, low on-chip memory architecture and one can very cheaply gain access to a number of parallel cores which is beneficial for HPC education and prototyping. The very low power nature of these architectures also means that there is potential for their use in future HPC machines, however there is a high barrier to entry in programming them due to the associated complexities and immaturity of supporting tools. 

In this paper we present our work on ePython, a subset of Python for the Epiphany and similar many-core co-processors. Due to the limited on-chip memory per core we have developed a new Python interpreter and this, combined with additional support for parallelism, has meant that novices can take advantage of Python to very quickly write parallel codes on the Epiphany and explore concepts of HPC using a smaller scale parallel machine. The high level nature of Python opens up new possibilities on the Epiphany, we examine a computationally intensive Gauss-Seidel code from the programmability and performance perspective, discuss running Python hybrid on both the host CPU and Epiphany, and interoperability between a full Python interpreter on the CPU and ePython on the Epiphany. The result of this work is support for developing Python on the Epiphany, which can be applied to other similar architectures, that the community have already started to adopt and use to explore concepts of parallelism and HPC.
\end{abstract}

\section{Introduction}
The Epiphany coprocessor \cite{epiphany-intro} has been developed by Adapteva and is packaged as a single chip component comprising of many of processing cores. The Parallella single board computer \cite{parallella} has been built around the Epiphany chip and combines it with a host ARM CPU. The most common Parallella ships for \$99 and provides a 16 core Epiphany III with a dual core host ARM. Whilst Adapteva sell this hardware as \emph{supercomputing for everybody} due to the low price point, actually programming this machine is time consuming and of considerable difficulty to novices. Being able to write high level parallel code and then experiment with parallelism in Python would be of great advantage for both education and fast prototyping, however the limited amount of per core memory (32KB) available on the Epiphany means that it is impractical to port an existing Python interpreter.

In this paper we present our work on supporting a subset of Python on high core count, low on-chip memory devices such as the Epiphany. By carefully designing an approach that takes advantage of the CPU during Python interpretation we can support the user writing Python parallel codes on the Epiphany. This is aimed at education and prototyping, where even novices can exploring the computational and parallel techniques that HPC programmers adopt on large scale modern supercomputers. After discussing the Epiphany and related work by the community in more detail in section 2, section 3 focuses in ePython itself. In this section we consider some of the general support for parallelism and running hybrid on both the CPU and Epiphany, we describe the architecture of ePython in more detail before examining the performance and programmability of an example Gauss-Seidel computational code. In section 4 we discuss interoperability with an existing Python interpreter running on the host and consider a heterogeneous example, where full Python on the host offloads computation to the Epiphany cores (in this case number sorting), before draw conclusions and discussing further work in section 5.

\section{Background}
\label{sec:bg}
Each Epiphany core comes with its own local, 32KB of on-chip memory and is interlinked to other cores via eMesh, a fast interconnect arranged in a 2D mesh with nearest neighbour direct connections. Each processor core in the Epiphany chip consists of a RISC CPU, high bandwidth local memory, DMA engine and network interface. Performance levels close to 2 GFLOP/s per core have been demonstrated \cite{epiphany} in signal processing kernels and this architecture is applicable to a wide range of uses which include many HPC codes. Adapteva have sold 64 core and 16 core Epiphany coprocessors but the architecture, which is designed for scalability, has been demonstrated with up to 4096 cores and a production 1024 core version (the Epiphany V) has been announced \cite{epiphanyV} which is compatible with previous generations and will be available in early 2017. The most common version in current circulation is the 16 core Epiphany III (32 GFLOP/s theoretical peak performance), with each core having 32KB of local memory distributed across four 8KB banks. It is expected that, as the architecture becomes more mature, then Epiphany coprocessors with a far greater numbers of cores will become more common place. In this paper we focus on the 16 core Epiphany III chip due to its ubiquity, granted this is more multi-core than many-core but the concepts and implementation discussed here can easily be applied to both single coprocessors with very many more cores (such as the announced 1024 core Epiphany V) and also multiple coprocessors interconnected via Adapteva's eMesh technology.

The Parallella single board computer comprises of a dual core ARM A9 CPU and 1GB host RAM, 32MB of which is reserved as shared memory which both the host and Epiphany coprocessor can directly access. With a price point starting at \$99 this has been developed as a low cost means of leveraging HPC as well as a technology demonstrator for the Epiphany itself. The Linux OS is shipped with the Parallella, like many accelerators the Epiphany is bare metal and has no operating system. Both the Parallella and Epiphany are open platforms, where all details such as the documentation, designs and entire source code are fully and freely available for anyone to use. Adapteva are keen for the community to develop tools, techniques and technologies which take advantage and make novel use of the Parallella and Epiphany.

Up until now the programmer of the Epiphany co-processor has only had a choice between writing their code in assembly or C. Commonly two executables are built, one launched on the host CPU and the other on the Epiphany cores. Shared memory is then used for communication and the marshaling of control flow between the host and device. An open source low level API has been developed which allows the programmer to call specific functions from their C code, such as installing interrupt timers and utilising the DMA engines of the cores. However even with the API, writing code for the Epiphany directly is fraught with difficulties, such as data consistency, different data alignments on the host and Epiphany and having to carefully select the appropriate linker script. As is common with these sorts of devices, the Epiphany has no IO connectivity, so novices can find it very difficult to debug their code where they would simply add a print statement to CPU code. 

There have been a number of technologies developed, with varying degrees of maturity, to support programming the Epiphany, such as COPRTHR\cite{cprthr}, a threaded MPI \cite{ross2016parallel} and OpenSHMEM \cite{ross2016implementing} but these  still require the programmer to address the limited amount of on-chip memory per core, to still worry about the tricky lower level details of the architecture and write their code in C. In addition to the on-chip memory there is an additional 32MB of CPU main memory available to be shared between the host and Epiphany. One might assume that the on-chip memory would be best viewed as cache, however because of a lack of hardware cache support and the slow transfer speeds from host memory it is preferable to maximise the use of on-chip memory and this is often viewed as ``main memory'' rather than cache, with the shared host memory avoided for code or program data. To put this in perspective libc is far larger than 32KB, so if one wishes to use common C functionality such as string handling, maths functionality or memory allocation then they have to locate libc in shared main memory and take the performance hit or implement ones own common C functionality but this adds to the complexity of user code and development time. In short the existing ways of programming the Epiphany are neither accessible to the novice nor do they encourage fast development and prototype experimentation with the architecture. The result is that, even though Adapteva sold the Parallella as democratising supercomputing, few people have the expertise to be able to write even a simple parallel code for the Epiphany and this has meant that many people have purchased the machine and then been unable to use it.

As well as the low cost of the Epiphany, one of the other major advantages is its energy efficiency, with the 16 core Epiphany III consuming less than 2 Watts of power \cite{E16G301datasheet}. Energy efficiency is generally seen as major challenge to reaching exa-scale and an architecture, for instance the Epiphany III, that can deliver 32 GFLOP/s theoretical peak performance for less than 2 Watts is of great interest for reaching our energy targets for exa-scale machines. The announced Epiphany V is predicted to deliver 75
GFLOPS/Watt \cite{epiphanyV}. Instead of large power hungry processors there is interest in utilising these very many low power, low memory cores but the programming challenges associated with taking advantage of these are significant. This technology is also of great interest for parallelism in embedded devices, such as smart phones, where both performance and energy efficiency are crucially important. Being able to effectively program and experiment with these sorts of technologies will be key to their future development and acceptance by the HPC community.

To enable both education and fast prototyping, support for writing Python code is an obvious activity. However the limited, 32KB, memory per core provides a significant challenge here. A full Python interpreter, such as CPython \cite{cpython} and associated libraries will nowhere near fit in a core's memory and in addition we also need to place the user's code and data in this same space. Other technologies such as Numba \cite{numba} exist which can JIT compile code for GPUs or CPUs, but again the compiled code and libraries are far too large to fit within an Epiphany core's memory. Even Python resources aimed at embedded devices require Megabytes rather than Kilobytes of memory to run. There has been work done supporting Python for the many-core Knights Landing architecture \cite{knlpython}, however KNL has vastly more memory than the Epiphany and as such this work is not directly applicable here.

\section{ePython}
\label{sec:epython}
The ePython interpreter has been developed to support a subset of Python on the Epiphany co-processor. The programmer writes their Python code and executes it via the ePython tool on the Parallella which then interprets the code and executes it on the Epiphany. It is mainly the imperative aspects of Python that are currently supported with a number of extensions for parallelism provided as importable modules. The intention of ePython has been to allow a complete novice to write their first simple parallel Python code and run this on the Epiphany in one minute or less.

\begin{lstlisting}[frame=lines,caption={Python parallel hello world code},label={lst:helloworld}]
from parallel import *

print "Hello world from core "+str(coreid())+" of "+str(numcores())
\end{lstlisting}

Listing \ref{lst:helloworld} illustrates a simple hello world Python code where each Epiphany core will display the message with its core ID and total number of Epiphany cores. The \emph{coreid} and \emph{numcores} functions are located in the \emph{parallel} module which is imported on line 1. One of the challenges of devices such as the Epiphany is that codes running on these can not interact with the user, not being able to display information to stdio is especially difficult when people are trying things out and debugging. ePython abstracts all of this from the programmer and supports user interaction (both output and input) from Python code running on any Epiphany core transparently.

\begin{lstlisting}[frame=lines,caption={Python point to point communication example},label={lst:p2p}]
from parallel import *

if coreid()==0:
  send(20, 1)
elif coreid()==1:
  print "Got value "+str(recv(0))+" from core 0"
\end{lstlisting}

Point to point message passing communication calls form a crucial part of many parallel codes, listing \ref{lst:p2p} illustrates a Python code to send the value \emph{20} from Epiphany core 0 at line 4 to core 1 which issues a receive (line 6) and displays the resulting value. Scalars or lists of integers, reals, booleans and strings can be communicated and currently all communications are blocking. In the case of communicating lists either the length of the list or an additional argument, if it is provided, will determine the number of elements to send or receive. In this example all other Epiphany cores (cores 2 to 15) will execute the conditional statements but effectively do nothing as their core ids do not match.
  
\begin{lstlisting}[frame=lines,caption={Python collective reduction example, finding the maximum \\random number held on an Epiphany core},label={lst:reduce}]
from parallel import *
from random import randint

a=reduce(randint(0,100), "max")
print "The highest random number is "+str(a)
\end{lstlisting}

Another important aspect of many parallel codes are collective communications, where all cores are involved in a specific communication to determine the result. Listing \ref{lst:reduce} illustrates a reduction in Python, where each Epiphany core will generate a random number between 0 and 100, this is then reduced (line 3) via the \emph{reduce} function call with the operator (in this case \emph{``max''}) as an additional argument. In this case the maximum random number generated by any core is selected and this is communicated to every other core which then stores it in variable \emph{a}. Each core then displays its value of \emph{a} (line 4) which will be identical on them all. Other operators, \emph{min}, \emph{sum} and \emph{prod} are also supported. ePython's parallel module also provides a \emph{bcast} function call which will broadcast a value from one specific root Epiphany core to every other core.

In all these examples the programmer executes their Python code in the standard way of saving it as a file and then providing the filename as an argument to the epython executable run from bash on the Parallella host. There are also a variety of other options which configure the specifics of how the code should be executed, for instance one can run on a subset of Epiphany cores and specify more advanced placement information such as utilising only a range of cores for their code. It was mentioned in section \ref{sec:bg} that in addition to each Epiphany core's 32KB of local memory there is also 32MB of slow shared host memory which can also be accessed. By default ePython will try to locate everything (the code and data) in the core's memory, but it is possible via command line options to override this and control which memory is used for what. 

\subsection{Architecture}
\label{sec:arch}
\begin{figure}
\begin{center}
\includegraphics[scale=0.65]{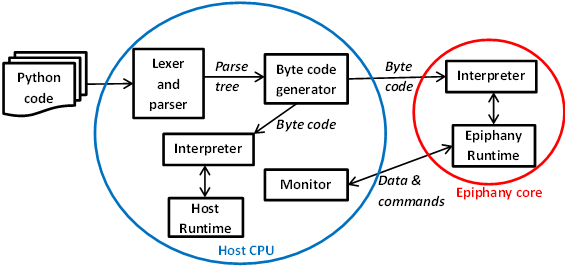}
\end{center}
\caption{ePython interpreter architecture, where byte code generation is done on the host and then transmitted to the Epiphany cores. The host also runs as a monitor to perform activities such as IO when directed by the Epiphany}
\label{fig:epyarch}
\end{figure}

Due to the limitations of the Epiphany coprocessor and compilers, the ePython interpreter is written in C with supporting Python modules that the programmer can import. Figure \ref{fig:epyarch} illustrates the overall architecture of ePython, due to the limited on core memory we aim to do as much of the preparation of user code as possible on the host CPU and only run the minimal amount on the Epiphany required to actually execute the code. A programmer's Python code is first preprocessed and then parsed on the host to form a parse tree, which is then operated upon to form byte code. For memory reasons this byte code is different to Python's standard byte code and optimised to be as small as possible. Each entry in the byte code starts with the ``command'' (such as conditional or variable assignment) which is 1 byte, the command will then have a number of different properties associated with it such as variable identifiers (each 2 bytes), operators (1 byte), constants (4 bytes), memory addresses (4 bytes) and relative locations in the byte code to jump to (2 bytes.) These commands are then appended sequentially to form the overall byte code and transferred to a portion of the shared memory, along with commands for activating the Epiphany cores.  

\begin{figure}
\begin{center}
\includegraphics[scale=0.76]{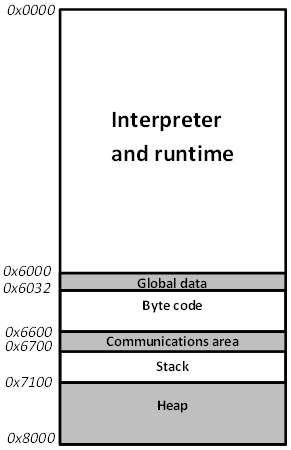}
\end{center}
\caption{ePython memory map for each Epiphany core's 32KB SRAM}
\label{fig:memmap}
\end{figure}

On the Epiphany itself the byte code is copied from shared memory into the local memory of each core. The interpreter then runs which reads in each command from the byte codes and processes it. This interpreter also maintains a stack area for each function call and heap for lists and strings. Larger codes or lists will transparently ``flow over'' to an area of the 32MB shared memory which is reserved for each core. This will have a performance impact but enables larger, more complex codes to execute. Figure \ref{fig:memmap} illustrates ePython's memory map for an Epiphany core, where the first 24KB is taken up with the interpreter and runtime. After this there is a section reserved for global variables (this is dynamic and in this example 50 bytes is allocated), next is the byte code itself (again dynamic and in this example 1.4KB), next is the communications area (fixed at 256 bytes) which supports message passing functionality. 1KB is then reserved for the stack and the remaining memory for the heap. 

The Epiphany runtime (figure \ref{fig:epyarch}) handles inter core communications where a target post-box, reserving memory for data from every other core, is allocated per core and is the 256 byte communications area of figure \ref{fig:memmap}. Messages are sent by writing to a specific post-box location in the target core's memory and then awaiting acknowledgement, messages are received by waiting on this local memory area. The actual data transfer is performed via the DMA engine of an Epiphany core and in addition to the message itself an additional byte (that wraps around) is also associated to provide versioning information which is crucial for data consistency. The runtime also provides IO functionality, the Epiphany itself can not directly perform IO but instead there is a monitor running inside a thread on the host. The monitor will wait on an area of shared memory for commands and data, such as displaying a message, and can write data back (for instance from user input) into shared memory that the Epiphany core will pick up. This monitor approach not only provides the illusion of IO directly from the Epiphany cores, but it is also possible to request the monitor to perform functionality such as string concatenation, complex mathematical functionality and handling error messages. This reduces the memory footprint on each Epiphany core as code to support these more complex functions is only required on the host.  

Figure \ref{fig:epyarch} also illustrates an interpreter and a runtime running on the host, the interpreter is designed to be very portable between architectures and it is only the runtime that is tied to the specifics of the technology. To go from one architecture to another, only a new runtime needs to be developed and a host runtime has been written so that, in addition to running code on the Epiphany cores, the programmer can also run their code on the Parallella's dual core ARM CPU at the same time. For memory reasons the interpreter itself is designed to support the minimum amount of functionality that ePython requires and the Python modules that come with ePython (such as \emph{parallel} and \emph{util}) contain higher level functionality which is imported by the programmer if required. During the parsing stage on the host, memory optimisation is also performed to omit unused functions that the programmer might have imported into their code but not used. In developing ePython's Epiphany runtime a major advantage is that this is an open platform with all the documentation, source code and even designs of the Epiphany open and freely available, such work on a proprietary/closed platform would have been far more difficult and time consuming.

\subsection{Virtual cores}
As mentioned in section \ref{sec:arch} ePython has been written in such a way that the interpreter is highly portable and it is only the runtime that is specific to an architecture. In addition to support for the Epiphany we have also developed variations of these that run on the Parallella's dual core ARM CPU. These are known as virtual cores, for all intents and purposes they look like fully fledged Epiphany cores to the Python programmer, operate and communicate as such, but instead each is running within a thread on the CPU. The number of virtual cores is defined as an argument to the epython executable and the core ids follow on sequentially from the Epiphany core IDs.

\begin{lstlisting}[frame=lines,caption={Python virtual cores example, which illustrates handling both \\physical (Epiphany) cores and virtual (threaded CPU) cores},label={lst:virtualcores}]
from parallel import ishost, isdevice, coreid

if ishost():
  print "Core number "+str(coreid())+" is a virtual core on the CPU"
elif isdevice():
  print "Core number "+str(coreid())+" is a physical Epiphany core"
\end{lstlisting}

Listing \ref{lst:virtualcores} illustrates Python code for determining which cores are physical Epiphany and which are virtual cores via the \emph{ishost} and \emph{isdevice} functions which are part of the \emph{parallel} module. For instance if run with all 16 Epiphany cores and 16 virtual cores, then cores 0 to 15 are reported as physical Epiphany cores and cores 16 to 31 virtual CPU cores. If run with 5 Epiphany cores and 10 virtual cores, then cores 0 to 4 are reported as physical Epiphany cores and 5 to 14 as virtual CPU cores.

The first reason for virtual cores is to allow users to experiment with writing parallel Python codes for more than 16 cores, there are later generations of the Epiphany chip with more cores planned so for educational purposes it is desirable that programmers can experiment with developing and running their code on more cores which virtual cores transparently supports. The other reason for virtual cores is to support writing heterogeneous, hybrid parallel codes, where the Epiphany acts as an accelerator and the host CPU offloads computationally intensive aspects of a code to this device which then sends the results back. A common use case would be to run with a number of Epiphany cores and one virtual core on the host CPU and, using the \emph{ishost} and \emph{isdevice} functions, diverge depending whether it is the Epiphany or CPU. 

\subsection{Parallel Gauss-Seidel example}
\label{sec:gauss}
Whilst the main purposes of ePython are education and fast prototyping, performance is a key consideration of parallelism and as such it is worth addressing the sort of performance that the Python programmer might expect from their code with ePython. The interpreted nature of ePython will inevitably result in a significant overhead compared to a similar C code running on the Epiphany, but the Python code will be far easier and quicker to develop and experiment with. Thus there is a trade-off between programmability and performance. 

Listing \ref{lst:gauss} illustrates a parallel code to solve Laplace's equation for diffusion in one dimension. This uses the successive over relaxation (SOR) variant of the Gauss Seidel iterative method for solving systems of linear equations. Lines 8 to 10 define the function \emph{fillInitialConditions} which sets the initial, boundary, conditions. Lines 12 to 22 compute the residual norm of the local data and then issue a reduction (line 18) to determine the global norm. If the initial norm (\emph{bnorm}) is provided as an argument to this function then the returned value is relative to this \emph{bnorm} (which we want for each iteration), otherwise the absolute residual norm is returned (which we want initially.) Lines 25 to 27 calculate the amount of local data per core and define the local list, \emph{data} to hold this (which is allocated in the heap memory and contains extra halo elements at the start and end.) At line 38 the loop for the iterations begins, for each iteration halo swapping (line 40 and 41) to left and right neighbours (where appropriate) is performed via the \emph{sendrecv} communication call and then, after calling the function to calculate the current relative residual, lines 47 to 49 contain the SOR Gauss-Seidel kernel. When the code completes the final residual and number of iterations is displayed by Epiphany core 0 (line 52.) 

It can be seen from listing \ref{lst:gauss} that this is a high level code, the programmer can clearly see the computational method and geometric decomposition adopted for parallelism. They are working with familiar HPC concepts, such as halo swapping and data distribution without having to worry about any of the lower level concerns that users of the Epiphany coprocessor with other programming technologies would need to address. It becomes very simple to experiment with computational and parallelism choices, such as the number of Epiphany cores to use, data decomposition, the global data size and over relaxation factor (line 6.) The lower level nature of other Epiphany programming technologies would often necessitate wide spread, significant, changes to the code when experimenting with these factors. 

\begin{lstlisting}[frame=lines,caption={Python Gauss Seidel with SOR solving Laplace's\\ equation for diffusion},label={lst:gauss}]
from parallel import *
from math import pow, sqrt

DATA_SIZE=1000
MAX_ITS=10000
W=1.3  # Overrelaxing factor (between 1 and 2)

def fillInitialConditions(local_data, local_size):
  if coreid()==0: local_data[0]=1.0
  if coreid()==numcores()-1: local_data[local_size+1]=10.0

def computeNorm(local_data, local_size, bnorm=none):
  tmpnorm=0.0
  i=1
  while i<=local_size:
    tmpnorm=tmpnorm+pow((local_data[i]*2-local_data[i-1]-local_data[i+1]), 2)
    i+=1
  tmpnorm=reduce(tmpnorm, "sum")
  if bnorm is none:
    return sqrt(tmpnorm)
  else:
    return sqrt(tmpnorm)/bnorm

# Work out the amount of local data
local_size=DATA_SIZE/numcores()
if local_size * numcores() != DATA_SIZE:
  if (coreid() < DATA_SIZE-local_size*numcores()): local_size=local_size+1

data=[0]*(local_size+2)

# Set the initial conditions
fillInitialConditions(data, local_size)

# Compute the initial absolute residual
bnorm=computeNorm(data, local_size)
norm=1.0
its=0
while norm >= 1e-3 and its < MAX_ITS:
  # Halo swap
  if (coreid() > 0): data[0]=sendrecv(data[1], coreid()-1)
  if (coreid() < numcores()-1): data[local_size+1]=sendrecv(data[local_size], coreid()+1)

  # Calculate current residual
  norm=computeNorm(data, local_size, bnorm)
  
  i=1
  while i<=local_size:
    data[i]=((1-W) * data[i]) + 0.5 * W * (data[i-1]+data[i+1])
    i+=1
  its+=1

if coreid()==0: print "Completed in "+str(its)+" iterations, RNorm="+str(norm)
\end{lstlisting}  

For comparison a version of this has been written in C, using Adapteva's own Epiphany API, compiled with GCC at optimisation level 3. At 266 lines this C code is far more verbose and much of this is taken up with lower level concerns such as data padding and transfer consistency. A performance evaluation with 1000 global elements, a relaxation factor of 1.3 and solving to a relative residual of 1e-3, has been performed with these versions and the results of this are illustrated in table \ref{tbl:performance}. The first table entry is running with ePython on all 16 Epiphany cores using default settings (all byte code and data in core memory.) The second entry in the table is still running on 16 Epiphany cores but the byte code and data have been located in shared memory rather than local Epiphany core memory (the ePython interpreter and runtime are still in Epiphany core memory.) Each core having to rely on shared memory for its byte code and data increases the runtime by 5 times and illustrates how significant an impact using the slow main memory is for performance. The third entry in the table illustrates ePython running as a virtual core on the host CPU only and when compared against the fourth entry, the common CPython implementation on the host CPU, it is clear that there is a significant overhead in the ePython interpreter compared to CPython. Even though there is this overhead, the serial CPython on the host is slower than ePython on 16 Epiphany cores. It can be seen that the versions in C, both running in serial on the CPU and in parallel on the Epiphany vastly outperform CPython and ePython. This is unsurprising, especially that the Epiphany C version is far faster than the ePython version, but this had to be hand crafted and took considerably longer to develop than the ePython code.

\begin{table}[h]
\caption{Runtime for different versions and configurations of Gauss Seidel SOR solver}
\label{tbl:performance}
\centering
\begin{tabular}{ | c | c | }
\hline
Runtime (s) \quad&\quad  \\
\hline			
9.61 \quad&\quad ePython on 16 Epiphany cores\\
52.04 \quad&\quad ePython all in shared memory\\
48.56 \quad&\quad ePython on host CPU only\\
14.71 \quad&\quad CPython on host CPU only\\
2.23 \quad&\quad C on ARM CPU only\\
1.01 \quad&\quad C on 16 Epiphany cores\\
\hline
\end{tabular}
\end{table}

\begin{figure}
\begin{center}
\includegraphics[scale=0.4]{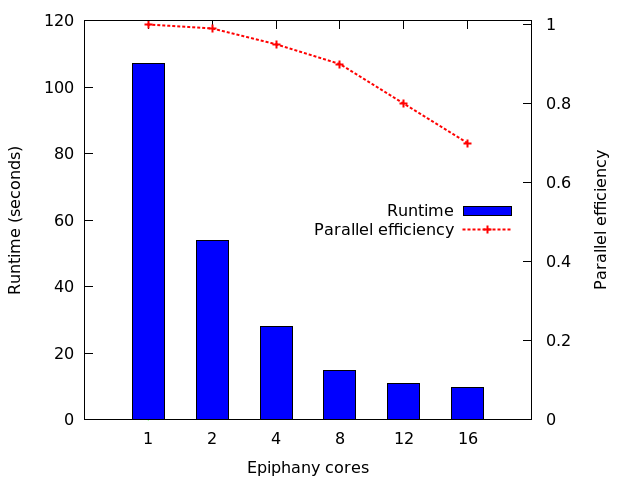}
\end{center}
\caption{Strong scaling Python runtime and parallel efficiency with varying Epiphany cores}
\label{fig:runtime_epcores}
\end{figure}

Figure \ref{fig:runtime_epcores} illustrates the runtime and parallel efficiency of the Python code running over different numbers of Epiphany cores. As one would expect as the number of Epiphany cores is increased the runtime drops and as we reach larger numbers of cores then we start to experience the limits of strong scaling, where the additional cost of communication outweighs the saving in computation. It was seen in table \ref{tbl:performance} that the performance of ePython is poor in comparison to C, but from figure \ref{fig:runtime_epcores} we can see that this is in the computational side of things as the code scales quite well based upon the parallel efficiency, even as the limits of strong scaling start to apply. This illustrates that there is potential for performance and scaling with ePython but there is work to be done optimising the execution speed. 

\section{Python Interoperability}
As mentioned in section \ref{sec:epython} ePython supports a subset of Python, as the community have picked up and started using ePython one of the requests amongst more advanced users has been to support full Python on the host running a common interpreter such as CPython. This allows for a rich Python code to be developed and then for specific aspects to be offloaded to the Epiphany and also to support existing Python codes where the programmer can experiment with offloading different aspects to the Epiphany. An extension has therefore been written, where ePython code runs on the Epiphany cores and full Python on the host CPU and the two can communicate. This is intended to help explore and prototype the offload model in more detail, where complex Python codes can offload aspects of their computation fairly easily using ePython. 

Two codes are written, a Python code to run on the Epiphany and second code to run on the host CPU. The host Python code imports the \emph{epython} module which provides all the communication calls (e.g. \emph{send}, \emph{recv} and \emph{reduce}) as they would appear in Python code on the Epiphany and these can be used to communicate with the code running on the Epiphany. The Python code running on the host is assigned a core ID one greater than the last Epiphany core ID. The \emph{epython} executable is then run with the code and also the \emph{fullpython} flag, the host Python code is run as normal with which ever Python interpreter the user prefers. Listing \ref{lst:cpuarraysort} illustrates the Python CPU code, generating random numbers and storing these in the \emph{data} list which is then sent (along with its length) to Epiphany core 0. The host Python code then blocks on the \emph{recv} call where it will receive the resulting sorted numbers back from Epiphany core 0 and then display them. A sketch of the Epiphany Python code is provided in listing \ref{lst:epythonsort}, where core 0 receives the length of data and data itself from the Python virtual core (id 16), all Epiphany cores call the \emph{parallel\_odd\_even\_sort} where core 0 will distribute the data amongst the other cores and these will sort it in parallel. Core 0 then sends the sorted numbers back to the host at line 10. The parallel odd even sort function is based upon one of the example Python codes at \cite{epythonsort}.

\begin{lstlisting}[frame=lines,caption={Python host CPU code for offloading parallel number sorting},label={lst:cpuarraysort}]
from random import randint
from epython import send, recv

data=[]
count=0
while count < 5000:
  data.append(randint(0,1000))
  count+=1

send(len(data), 0)
send(data, 0, len(data))
data=recv(0, len(data))

count=0
while count < 5000:
  print data[count]
  count+=1
\end{lstlisting}

\begin{lstlisting}[frame=lines,caption={Sketch of Python code running on Epiphany cores for\\ parallel number sorting},label={lst:epythonsort}]
from parallel import coreid, recv, send

data=[]
datalen=0
if coreid==0:
  datalen=recv(16)
  data=recv(16, datalen)
parallel_odd_even_sort(data)
if coreid==0:
  send(data, 16, len(data))
\end{lstlisting}

Table \ref{tbl:isperformance} illustrates the runtime of ePython on the Epiphany and CPython on the host sorting 5000 numbers via the odd even sort method. The first entry illustrates the hybrid code on 16 Epiphany cores and the host CPU (running CPython) compared to the second entry which is the runtime of a serial version in CPython on the host only. Unlike the Gauss-Seidel example, it can be seen here that using serial CPython for sorting the numbers is slightly faster than sorting on the Epiphany cores via the hybrid offload approach of ePython+CPython. There are a number of reasons for this, firstly CPython is far more mature and will contain optimisations which ePython does not yet implement. Secondly shared memory between the host and Epiphany is slow (as discussed in section \ref{sec:gauss}) and data must negotiate this in the hybrid case to communicate with the Epiphany cores. Additionally the hybrid code requires core 0 to distribute its received data amongst the other cores and then gather this back up afterwards before sending it back to the host and this adds extra overhead. 

\begin{table}[h]
\caption{Runtime comparison for number sort}
\label{tbl:isperformance}
\centering
\begin{tabular}{ | c | c | }
\hline
Runtime (s) \quad&\quad  \\
\hline			
43.83 \quad&\quad Hybrid ePython and CPython\\
37.59 \quad&\quad CPython on host only\\
\hline
\end{tabular}
\end{table}

\section{As an educational tool}
The main driving force behind the development of ePython is using it as an educational tool. Not everyone has access to a parallel machine that they can play around with and nor do they have the in-depth programming experience to be able to write parallel codes in more traditional approaches such as C+MPI. By abstracting the user from the lower level details they can concentrate on the higher level concerns of their code both from a computation and parallel point of view. ePython provides a limited number of parallel message passing calls in comparison with a technology such as MPI, but as we have seen these are sufficient for a large number of codes and teach the underlying concepts.

Python codes illustrating a variety of different parallel patterns such as the geometric decomposition of our Gauss-Seidel example, pipeline, master worker and parallel divide and conquer have been developed and are available in the ePython repository as examples. Additionally a series of blog articles have been published on the Parallella website, initially aimed at empowering novices to write a simple parallel code in less than 1 minute these have then looked at aspects of parallel messaging and different approaches for parallelising problems. Common HPC computational codes are also introduced both in the blog articles and ePython examples. Through all of these there are some very interesting behaviours exhibited which people can experiment with, we have seen one example of this in section \ref{sec:gauss} where the limits of strong scaling mean that as a greater number of cores is used then there is a law of diminishing returns. With smaller global data sizes people can actually see an optimum number of Epiphany cores and then a decrease in performance when they go past this, all due to strong scaling.

The community have contributed a number of examples such as Mandlebrot, a Monte Carlo method (dartboard to generate PI) and sorting algorithms. There are other tools, such as the Jupyter Notebook \cite{ipython} which community members have integrated with ePython for interactive programming of examples to aid education in parallelism. Very many of the ideas and concepts that people play with in Python on the Epiphany can then be scaled up and apply to writing parallel codes for large scale supercomputers. 

The Parallella is not the only technology that supports HPC education, for instance for a comparable price it is possible to purchase a number of Raspberry PIs and this has been done in Wee Archie \cite{weearchie} which is made up of 18 Raspberry PIs. However in this approach one not only needs to buy the boards but also to purchase all the supporting hardware such as the network switches and cables. They also need the expertise and time to set both the hardware and software environment up so that these can communicate before being able to write or experiment with any parallel code. Networking on the Raspberry PI is implemented on top of USB and driven in software, therefore it is notoriously slow and as such running codes over multiple nodes often results in worse performance than a single node. With the Parallella and ePython there is no additional setup or building costs and the Epiphany inter-core eMesh connection technology is such that inter-core communications will perform much better and closer or even better to what one would expect from large scale HPC machines.      

\section{Conclusions}
In this paper we have introduced ePython, an implementation of a subset of Python for the Epiphany co-processor. Whilst architectures such as the Epiphany have great potential, both from an educational (due to low cost) and future HPC (due to very low power) perspective these chips are very difficult to program for a variety of reasons. Due to the high level nature of Python, our approach allows users to quickly develop parallel codes and experiment with a variety of different considerations and approaches. Combined with the low cost nature of the Parallella, ePython could become a key enabler for making supercomputing accessible to all. There is also benefit from a prototyping point of view, it is true that eventually one would want to port their code to C for performance reasons but initially they can very easily and quickly develop and experiment using Python that can then inform and drive the porting into C later on. The central ideas behind ePython are not unique to the Epiphany, our work can be applied to other high core count, low memory architectures and Python could act as an approach for quickly testing and evaluating further developments in this area in the future.

There is plenty of further work for ePython especially because the 1024 core Epiphany V, which will be available in early 2017, will likely attract a lot of interest and new users to the technology. Being able to quickly get started and write high level parallel code in Python will be a major advantage and an obvious first step will be to further optimise the interpreter. By using techniques such as JIT compiling and static optimisation it should be possible to significantly increase the performance of ePython, to make it more competitive with other Python interpreters and much closer to C on these architectures. Another obvious extension is on the parallel side of things, non-blocking messages would be useful and also support for shared memory communication would be beneficial and support education in writing shared memory parallel codes. When combining ePython on the Epiphany with a full Python interpreter, such as CPython, on the host it is not ideal to require two separate codes nor communicate via the message passing function calls. Instead further abstraction, where ePython code is contained as functions in the host code (probably decorated with some annotation) and the programmer could interact with this RPC style, by calling these functions on the host, and the underlying \emph{epython} module would handle all the communications to and from the Epiphany transparently. 

These further developments to ePython combined with Adapteva'a future roadmap for the Epiphany architecture will provide a strong foundation for prototyping and experimenting with parallelism and parallel algorithms in Python over thousands of low power cores.

ePython is open source, licenced under BSD and available for download at \\ \textbf{https://github.com/mesham/epython}
\bstctlcite{BSTcontrol}
\bibliographystyle{IEEEtran}
\bibliography{IEEEfull,document}
\end{document}